%% file: main.tex
\documentclass[sigconf]{acmart} 
\usepackage{tabularx,booktabs,threeparttable,array}
\usepackage{pifont} 
\usepackage{pifont}                     
\newcommand{\cmark}{\ding{51}}          
\usepackage{multicol,multirow}
\usepackage[table]{xcolor}
\usepackage{lstlinebgrd}
\newcolumntype{W}{>{\raggedright\arraybackslash}X}   
\newcolumntype{C}{>{\centering\arraybackslash}X}     

\usepackage{listings}
\usepackage{xcolor} 
\definecolor{codegreen}{rgb}{0,0.6,0}
\definecolor{codegray}{rgb}{0.5,0.5,0.5}
\definecolor{codepurple}{rgb}{0.58,0.3,0.82}
\definecolor{backcolour}{rgb}{0.95,0.95,0.92}

\newcommand{\chref}[2]{%
  \setlength{\fboxrule}{0.4pt}%
  \setlength{\fboxsep}{1pt}%
  \fcolorbox{red}{white}{\href{#1}{#2}}%
}

\lstdefinestyle{mystyle}{
    language=python,
    backgroundcolor=\color{backcolour},   
    commentstyle=\color{codegreen},
    keywordstyle=\color{magenta},
    numberstyle=\tiny\color{codegray},
    stringstyle=\color{codepurple},
    basicstyle=\ttfamily,
    breaklines=true,
    captionpos=b,                    
    keepspaces=true,                 
    numbers=left,                    
    numbersep=5pt,                  
    showspaces=false,                
    showstringspaces=false,
    showtabs=false,                  
    tabsize=2,
    frame=tb,                 
    rulecolor=\color{gray},   
    framerule=0.9pt,
    escapeinside={(*}{*)}
}

\usepackage[linesnumbered,ruled,vlined]{algorithm2e}

\newcommand{\fullbg}[2]{
  \begingroup\setlength{\fboxsep}{0pt}%
  \rlap{\colorbox{#1}{\makebox[\linewidth]{\strut}}}#2\endgroup}

\definecolor{headbg}{gray}{0.93}
  
\begin{document}

\title{Specification and Detection of LLM Code Smells}

\author{Brahim MAHMOUDI}
\authornote{The first two authors contributed equally to this work and share first authorship.}
\orcid{0009-0008-3007-7080}
\affiliation{%
  \institution{École de technologie supérieure}
  \city{Montréal}
  \state{Québec}
  \country{Canada}
}
\author{Zacharie CHENAIL-LARCHER}
\orcid{0009-0002-7464-6659}
\affiliation{%
  \institution{École de technologie supérieure}
  \city{Montréal}
  \state{Québec}
  \country{Canada}
}
\authornotemark[1]

\author{Naouel MOHA}
\orcid{0000-0001-9252-9937}
\affiliation{%
  \institution{École de technologie supérieure}
  \city{Montréal}
  \state{Québec}
  \country{Canada}
}

\author{Quentin STIÉVENART}
\orcid{0000-0001-9985-9808}
\affiliation{%
  \institution{Université du Québec à Montréal}
  \city{Montréal}
  \state{Québec}
  \country{Canada}
}

\author{Florent AVELLANEDA}
\orcid{0000-0003-1030-5388}
\affiliation{%
  \institution{Université du Québec à Montréal}
  \city{Montréal}
  \state{Québec}
  \country{Canada}
}


\input{Sections/abstract}

\ccsdesc[500]{Software and its engineering~Software reliability}
\ccsdesc[300]{Software and its engineering~Software maintenance and evolution}
\ccsdesc[300]{Software and its engineering~Software design engineering}
\ccsdesc[100]{Computing methodologies~Natural language processing}

\keywords{Large Language Models, LLMs, Code Smells, LLM Integration}


\maketitle

\section{Introduction}

\input{Sections/introduction}
\section{Related Work}

\input{Sections/related_works}

\section{Catalog of LLM Code Smells}

\input{Sections/identify_llm_code_smells}


\section{Methodology}
\input{Sections/Methodology}

\section{Empirical Validation}

\input{Sections/Empirical_Validation}

\section{Future Plans}

\input{Sections/Future_plans}

\section{Conclusion}

\input{Sections/conclusion}

\bibliographystyle{ACM-Reference-Format}
\bibliography{bibli}

\appendix

\end{document}

%% file: Sections/abstract.tex
\begin{abstract}

Large Language Models (LLMs) have gained massive popularity in recent years and are increasingly integrated into software systems for diverse purposes. However, poorly integrating them in source code may undermine software system quality. Yet, to our knowledge, there is no formal catalog of code smells specific to coding practices for LLM inference. In this paper, we introduce the concept of LLM code smells and formalize five recurrent problematic coding practices related to LLM inference in software systems, based on relevant literature. We extend the detection tool \textit{SpecDetect4AI} to cover the newly defined LLM code smells and use it to validate their prevalence in a dataset of 200 open-source LLM systems. Our results show that LLM code smells affect 60.50\% of the analyzed systems, with a detection precision of 86.06\%.
\end{abstract}

%% file: Sections/introduction.tex
In recent years, Large Language Models (LLMs) have revolutionized the way information is processed and have gained increasing importance in everyday life, with the number of LLM-related publications multiplying year after year ~\cite{xia2025analyzing16193llmpapers}.  They are also integrated into a growing number of software systems ~\cite{shao2025llmscorrectlyintegratedsoftware}.

However, LLMs are not always reliable ~\cite{khatun2024reliability}. Their performance and behavior can vary significantly depending on how they are used ~\cite{YANG2025113503}. To ensure the reliability and maintainability of \textbf{LLM-integrating systems}, software systems that invoke LLMs at runtime and use their outputs in program logic, it is essential to integrate them properly at both the architectural and source code levels~\cite{bucaioni2025functionalsoftwarereferencearchitecture}. Therefore, as previously done for general ~\cite{10.5555/311424} and machine learning-specific practices \cite{zhang2022codesmellsmachinelearning}, establishing coding guidelines, including the formalization of code smells, is essential for LLM-integrating systems.

While prior studies have defined taxonomies of general~\cite{shao2025llmscorrectlyintegratedsoftware} and prompt-related defects in these systems~\cite{ tian2025taxonomypromptdefectsllm}, to our knowledge, there is no formal concept addressing code-specific poor practices for the integration of LLM inference in software systems. 



In this paper, we introduce the concept of LLM code smells and specify five concrete cases identified from relevant literature. We define them as recurring source code patterns tied to LLM inference that, while not directly causing bugs or failures, undermine maintainability, reliability, performance, or robustness. 


Our key contributions are: (i) a catalog of five concrete LLM code smells; (ii) \chref{https://github.com/Brahim-Mahmoudi/SpecDetect4LLM_ICSE}{\textit{SpecDetect4LLM}}, a dedicated static detection tool; and (iii) an empirical study on 200 Python open-source systems to assess their prevalence and validate their relevance.

%% file: Sections/related_works.tex
% ...



The concept of \textbf{code smells} is well established in software engineering. It can be defined as a recurring, low-level coding practice that degrades software quality, while not representing an explicit bug ~\cite{10.5555/311424}. Over time, researchers have extended the concept by developing catalogs and taxonomies for specific application domains, such as for Android applications ~\cite{Carvalho2019AndroidSmells} and Machine Learning (ML) ~\cite{zhang2022codesmellsmachinelearning}. We follow a similar approach for LLM-based systems.

The closest existing work to our study is the taxonomy of code defects for LLM-based autonomous agents introduced by \textbf{Ning et al.}~\cite{ning2024definingdetectingdefectslarge}. While some of their defects, such as \textit{LLM API-related Defect}, touch similar aspects to our code smells, both concepts differ significantly in scope, granularity, and effects. We focus on LLM inference calls in system source code, which occur both in simple integrations and in their agent systems~\cite{ke2025surveyfrontiersllmreasoning}. Their defects are also broader and more general, adopting a higher-level view, while we precisely pinpoint poor coding practices. Finally, their defects directly cause failures and malfunctions, whereas LLM code smells harm the maintainability, reliability, performance and robustness. \textbf{Shao et al.} ~\cite{shao2025llmscorrectlyintegratedsoftware} introduce a taxonomy of defects for LLM-enabled and RAG systems. Their defects are fundamentally different from our code smells, as they describe higher-level symptoms not explicitly tied to recurrent coding practices.


Other related work can be organized into four categories: (i) \textbf{taxonomies of failures} in LLM-integrating systems that emphasize high-level symptoms rather than code-level practices~\cite{11081716,cemri2025multiagentllmsystemsfail,le-jeune-etal-2025-realharm}; (ii) \textbf{prompt-smell taxonomies} that are complementary to our study, but operate on prompts, not code \cite{ronanki2024promptsmellsomenundesirable,tian2025taxonomypromptdefectsllm}; (iii) \textbf{anti-patterns} for LLM inference \textbf{benchmarking} rather than integration~\cite{agrawal2025evaluatingperformancellminference}; and (iiii) \textbf{analyses of defects in LLM-generated code}, whereas we study human-written source code that integrate LLMs~\cite{zhuo2025identifyingmitigatingapimisuse,esfahani2024understandingdefectsgeneratedcodes}.

Unlike prior works that catalog agent defects, prompt smells, or runtime failures, we define \textbf{LLM code smells} at the source-code level, \textbf{create a detailed catalog}, \textbf{extend} \textit{SpecDetect4AI} \textbf{into \chref{https://github.com/Brahim-Mahmoudi/SpecDetect4LLM_ICSE}{\textit{SpecDetect4LLM}} to detect} them, and \textbf{conduct a prevalence assessment} across diverse open-source systems.

%% file: Sections/identify_llm_code_smells.tex
This section covers our catalog of five LLM code smells. Table~\ref{tab:defects-llm} summarizes their sources, effects on software quality, and prevalence. While we focus on Python code, the concepts also apply across languages and use cases. Due to space constraints, we include extended explanations/definitions in our replication package~\cite{Replication_package}.

\newcolumntype{E}{>{\hsize=.25\hsize\centering\arraybackslash}X}
\begingroup
\captionsetup{aboveskip=1pt,belowskip=1pt} 
\renewcommand{\arraystretch}{0.82}          
\setlength{\extrarowheight}{0pt}

\setlength{\aboverulesep}{0.2ex}
\setlength{\belowrulesep}{0.2ex}
\setlength{\cmidrulesep}{0.3ex}

\renewcommand{\cmark}{\smash{\raisebox{0.06ex}{\scalebox{0.71}{\ding{51}}}}}

\renewcommand{\tabcolsep}{5pt} 

\makeatletter
\g@addto@macro\TPT@defaults{\setlength{\itemsep}{0pt}\setlength{\parsep}{0pt}}
\makeatother

\begin{table*}[t]
\centering
\footnotesize
\begin{threeparttable}
\caption{Overview of LLM code smells: sources, effects, prevalence, and detection precision.}

\label{tab:defects-llm}
\renewcommand{\arraystretch}{1}

\begin{tabularx}{\textwidth}{@{} l *{3}{C} *{4}{E} c c r r r @{}}
\toprule
\multirow{2}{*}{\textbf{Code Smell}} &
\multicolumn{3}{c}{\textbf{Sources}} &
\multicolumn{4}{c}{\textbf{Effects}} &
\multicolumn{2}{c}{\textbf{Prevalence}} &
\multicolumn{3}{c}{\textbf{Precision}} \\
\cmidrule(lr){2-4}\cmidrule(lr){5-8}\cmidrule(lr){9-10}\cmidrule(lr){11-13}
& \textbf{Lit.} & \textbf{Grey} & \textbf{Emp.} & \textbf{RO} & \textbf{P} & \textbf{M} & \textbf{R} & \textbf{\%} & \textbf{\#\#} & \textbf{TP} & \textbf{FP} & \textbf{P(\%)} \\

\midrule
\textbf{UMM}  & \cite{HanEtAl2025TokenBudgetAware, chen2025adaptivelyrobustllminference} & \cite{OpenAIDocs2025,AnthropicDocs2025,GoogleGeminiDocs2025,AzureOpenAIDocs2025,BigQuery2025Quotas, aws2025timeouts, openai_python2025}
 &  & \cmark  & \cmark  & \cmark &  &  38.00\% & 76/200 & 19 & 3 & 86.36 \\
 \textbf{NMVP} & \cite{WilsonEtAl2014BestPractices,VenturiniEtAl2023IDepended,ReyesEtAl2024BUMP, Morishige/EnsureReprod} & \cite{AnthropicDocs2025,Microsoft2025FoundationLifecycle,HFTransformersDocs2025,OpenRouterHome2025,GoogleGeminiDocs2025} &  &  &  & \cmark  & \cmark &  36.00\% & 72/200 & 22 & 1 & 95.65 \\
 \textbf{NSM}  & \cite{jeong2025messagegenerationuserpreferences, Neumann_2025} & \cite{PromptHub2025SystemMessages,StackOverflow2023SystemRoleUseCase,OpenAIDocs2025,HFTransformersDocs2025} &  &  &  & \cmark  & \cmark &  34.50\% & 69/200 & 18 & 3 & 85.71 \\
\textbf{NSO}  & \cite{NeedStructuredOutput/10.1145/3613905.3650756} & \cite{Kharitonov2024EnforcingJSON,DeveloperService2025PydanticLLM,WymanBarber2024ValidateOutputs,Modelmetry2024JSONSchema,OpenAIDocs2025,AzureOpenAIDocs2025, OpenAIStructuredOutput}
 & \cite{LangChain2023Issue3709,ArenaAI2025StructuredLogprobs} & \cmark &  &  & \cmark &  40.50\% & 81/200 & 16 & 4 &80.00\\
 \textbf{TNES} & \cite{MinhEtAl2025MinpSampling,MontandonEtAl2025DABC} & \cite{AnthropicDocs2025,GoogleGeminiDocs2025,OpenAIDocs2025,VellumTemperature2025,HFTransformersDocs2025, OllamaModelfile2025} &  &  &  & \cmark  & \cmark  &  36.50\% & 73/200 & 19 & 4 & 82.61 \\
 
\bottomrule
\end{tabularx}

\fontsize{6.5}{6}\selectfont
\begin{tablenotes}
\item[*] \textit{Sources} -Literature: peer-reviewed literature; Grey: grey literature (official docs, tech reports, blogs); and Empirical: primary empirical data (e.g., commits, repos).
\item[*] \textit{Effects} - RO: Robustness (error-prone); P: Performance (cost, latency, memory); M: Maintainability (reproducibility, portability); R: Reliability (response quality, consistency, stability over time).
\item[*] \textit{Prevalence} - \fontsize{7}{7}\% : affected systems; \#\# = number of affected system by the code smell/total number of systems .
\item[*] \textit{Precision} - TP: True Positive; FP: False Positive; P: Precision.
\end{tablenotes}
\end{threeparttable}
\end{table*}

\endgroup

\subsection{\noindent\textbf{Unbounded Max Metrics (UMM)}}


\noindent\textbf{Context:} Hosted LLM APIs (e.g., OpenAI, Anthropic) expose finite token windows, per-request output caps, and rate limits. Pipelines that ignore these constraints, or omit their own limits on concurrency and response size, are prone to throttling and partial outputs.

\noindent\textbf{Problem:} Leaving token budgets, timeouts, and retries unbounded undermines system robustness and performance. Unspecified token budgets may yield outputs that exceed context limits (truncating mid-structure or return 400), overload downstream parsers, or inflate token costs and memory usage~\cite{chen2025adaptivelyrobustllminference}. Not specifying timeouts and retry limits can cause requests to hang indefinitely, leading to unpredictable latency, rising costs, and reduced throughput as resources are tied up by long-running calls~\cite{aws2025timeouts}. Not specifying these values also hinders maintainability, as defaults may change over time and tracking settings is essential for reproducibility.

\noindent\textbf{Solution:} Always bound and adjust the \textit{max\_output\_tokens}, \textit{timeout}, and \textit{max\_retries} parameters~\cite{openai_python2025}. Monitoring the number of input tokens is also recommended.


\noindent \textbf{Example: }In Listing~\ref{fig:UMM}, the red version leaves metrics unbounded, whereas the green enforces bounded metrics.

\begin{lstlisting}[style=mystyle,basicstyle=\ttfamily\fontsize{5.5}{6}\selectfont,language=python,alsoletter=_,caption=Unbounded Max Metrics (UMM),keywordstyle=\color{magenta},morekeywords={OpenAI,OpenAIError,RateLimitError,responses,create,with_options},deletekeywords={and,not,set,in},label=fig:UMM,linebackgroundcolor={
    \ifnum\value{lstnumber}=1\relax \color{red!15}\fi
    \ifnum\value{lstnumber}=2\relax \color{red!15}\fi
    \ifnum\value{lstnumber}=3\relax \color{red!15}\fi
    \ifnum\value{lstnumber}=4\relax \color{green!15}\fi
    \ifnum\value{lstnumber}=5\relax \color{green!15}\fi
    \ifnum\value{lstnumber}=6\relax \color{green!15}\fi
    \ifnum\value{lstnumber}=7\relax \color{green!15}\fi
    \ifnum\value{lstnumber}=8\relax \color{green!15}\fi
  }]
- client = OpenAI()
- response = client.responses.create(model=
    "gpt-4o-2024-11-20", input=prompt)
+ client = OpenAI(timeout=20, max_retries=3)
+ response = client.responses.create(
+    model="gpt-4o-2024-11-20", input=prompt, max_output_tokens=256)       
\end{lstlisting}

\subsection{\noindent\textbf{No Model Version Pinning (NMVP)}}


\noindent\textbf{Context:} In LLM APIs, runtimes, and hubs (e.g. OpenAI, Ollama, Hugging Face), models can be called via moving aliases (e.g., \textit{gpt-4o}) or immutable versions/snapshots (e.g., \textit{gpt-4o-2024-11-20}). Aliases may advance as providers update models~\cite{Microsoft2025FoundationLifecycle}.

\noindent\textbf{Problem:} Using only a provider alias removes explicit versioning, so weights, prompts, and safety filters can change without notice and shift behavior~\cite{Morishige/EnsureReprod}. It reduces maintainability by eroding traceability and reproducibility. Runs cannot be tied to a stable model build and portability across environments degrades as the same alias may yields different behavior.

\noindent\textbf{Solution:} Always specify an immutable identifier and record it with other run metadata. Update versions via change control~\cite{HFTransformersDocs2025, OpenAIDocs2025, WilsonEtAl2014BestPractices}.


\noindent\textbf{Example:} In Listing~\ref{fig:NMVP}, the red version uses a moving alias, whereas the green version pins an immutable version/snapshot to ensure reproducibility and traceability.

\begin{lstlisting}[style=mystyle,basicstyle=\ttfamily\fontsize{6}{7}\selectfont, language=python,alsoletter=_,caption=No Model Version Pinning (NMVP),keywordstyle=\color{magenta},morekeywords={openai, create, action,isMLMethodCall,hasExplicitHyperparameters},deletekeywords={and,not,set,in}, label=fig:NMVP,linebackgroundcolor={
    \ifnum\value{lstnumber}=1\relax \color{backcolour}\fi
    \ifnum\value{lstnumber}=2\relax \color{red!15}\fi
    \ifnum\value{lstnumber}=3\relax \color{green!15}\fi
  }]
response = openai.chat.completions.create(
  - model="gpt-4o", messages=messages )
  + model="gpt-4o-2024-11-20", messages=messages)
\end{lstlisting}

\subsection{\noindent\textbf{No System Message (NSM)}}


\noindent\textbf{Context:} In role-based chat APIs and runtimes, the system message sets global behavior, constraints, and tone for the assistant.

\noindent\textbf{Problem:} Without a system message, the model lacks high-level guidance, which reduces consistency and adherence to constraints. Outputs become more generic and harder to control. This degrades reliability, as additional iterations or longer prompts are often required to achieve adequate results~\cite{PromptHub2025SystemMessages}.

\noindent\textbf{Solution:} Always include a clear system message that defines roles, goals, and constraints. Keep task specifics in the user message~\cite{jeong2025messagegenerationuserpreferences}. 

\noindent\textbf{Example:} In Listing~\ref{fig:NSM}, the red version omits a \textit{system} message, whereas the green version adds a concise system instruction to anchor behavior, improve consistency and response quality.

\begin{lstlisting}[style=mystyle,basicstyle=\ttfamily\fontsize{6}{7}\selectfont, language=python,alsoletter=_,caption=No System Message (NSM),keywordstyle=\color{magenta},morekeywords={openai, create},deletekeywords={and,not,set,in}, label=fig:NSM,linebackgroundcolor={
    \ifnum\value{lstnumber}=1\relax \color{backcolour}\fi
    \ifnum\value{lstnumber}=2\relax \color{backcolour}\fi
    \ifnum\value{lstnumber}=3\relax \color{red!15}\fi
    \ifnum\value{lstnumber}=4\relax \color{red!15}\fi
    \ifnum\value{lstnumber}=5\relax \color{green!15}\fi
    \ifnum\value{lstnumber}=6\relax \color{green!15}\fi
    \ifnum\value{lstnumber}=7\relax \color{green!15}\fi
    \ifnum\value{lstnumber}=8\relax \color{green!15}\fi
  }]
response = openai.chat.completions.create(
  model="gpt-4o-2024-11-20",
- messages=[{"role": "user", "content": "Explain 
recursion with an example"}])
+ messages=[{"role": "system", "content": "You are a
Computer Science tutor. Answer clearly."},{"role":"user",
"content": "Explain recursion with an example"}])
\end{lstlisting}

\subsection{\noindent\textbf{No Structured Output (NSO)}}


\noindent\textbf{Context:} LLM-integrating systems often expect typed fields (e.g., JSON) but rely on LLM inference outputs which may not respect the format and produce raw free-form text. This smell applies when the output is later parsed, indexed, or executed assuming structure.

\noindent\textbf{Problem:} Without an enforced output schema, the system may receive free-form text where structured fields are expected~\cite{NeedStructuredOutput/10.1145/3613905.3650756}. This increases error-proneness via schema drift, missing or renamed fields, type mismatches, and silent truncation that passes as success, breaking parsers and downstream steps~\cite{NeedStructuredOutput/10.1145/3613905.3650756}. It degrades reliability as runs become inconsistent, data stores accumulate erroneous values, and execution/storage/display paths face injection~\cite{DeveloperService2025PydanticLLM}.

\noindent\textbf{Solution:} Enforce structured output at the API boundary. With OpenAI, declare a JSON Schema via \textit{response\_format} (chat completions) or \textit{text.format} (responses). With the Python SDK, you may instead bind the format directly to classes~\cite{OpenAIStructuredOutput}. Always validate results to handle refusals or other errors~\cite{Kharitonov2024EnforcingJSON, WymanBarber2024ValidateOutputs, OpenAIStructuredOutput}.



\noindent\textbf{Example:} In Listing~\ref{fig:NSO}, the red version consumes free-form text, whereas the green version enables strict parsing and safer downstream use by enforcing a JSON schema via \texttt{response\_format}.

\begin{lstlisting}[style=mystyle,basicstyle=\ttfamily\fontsize{5.5}{6.1}\selectfont, language=python,alsoletter=_,caption=No Structured Output (NSO),keywordstyle=\color{magenta},morekeywords={openai,create,ChatCompletion},deletekeywords={and,not,set,in},label=fig:NSO,linebackgroundcolor={
    \ifnum\value{lstnumber}=1\relax \color{backcolour}\fi
    \ifnum\value{lstnumber}=2\relax \color{red!15}\fi
    \ifnum\value{lstnumber}=3\relax \color{red!15}\fi
    \ifnum\value{lstnumber}=4\relax \color{green!15}\fi
    \ifnum\value{lstnumber}=5\relax \color{green!15}\fi
        \ifnum\value{lstnumber}=6\relax \color{green!15}\fi
        \ifnum\value{lstnumber}=6\relax \color{green!15}\fi
        \ifnum\value{lstnumber}=7\relax \color{green!15}\fi
        \ifnum\value{lstnumber}=8\relax \color{green!15}\fi
        \ifnum\value{lstnumber}=9\relax \color{green!15}\fi
  }]
# Define a JSON schema e.g. result_schema = ...
- response = openai.chat.completions.create(
-     model="gpt-4o-2024-11-20", messages=messages)
+ response = openai.chat.completions.create(
+     model="gpt-4o-2024-11-20", response_format={ "type": "json_schema",
+         "json_schema": {"name": "Result", "schema": 
+         result_schema}}, messages=messages)
\end{lstlisting}

\subsection{\noindent\textbf{LLM Temperature Not Explicitly Set (TNES)}}


\noindent\textbf{Context:} In LLM APIs, SDKs, and runtimes (e.g., OpenAI, Hugging Face, Ollama), temperature controls sampling randomness~\cite{VellumTemperature2025}. 

\noindent\textbf{Problem:} Relying on an implicit temperature reduces maintainability and reliability. Defaults temperature differ across providers/models~\cite{OpenAIDocs2025, GoogleGeminiDocs2025, OllamaModelfile2025} and may change over time~\cite{MontandonEtAl2025DABC}, which harms reproducibility, portability and can silently alter behavior. 

\noindent\textbf{Solution:} Always specify explicitly the temperature and document it. Tune by task: low (0–0.3) for precise, repeatable automation; higher (0.7–1.0) for creative generation; avoid extremes~\cite{MinhEtAl2025MinpSampling}.


\noindent\textbf{Example:} In Listing~\ref{fig:TNES}, the red version omits temperature, while the green version makes it explicit.

\begin{lstlisting}[style=mystyle,basicstyle=\ttfamily\fontsize{6}{7}\selectfont, language=python,alsoletter=_,caption=LLM Temperature Not Explicitly Set (TNES),keywordstyle=\color{magenta},morekeywords={openai, create, action,isMLMethodCall,hasExplicitHyperparameters},deletekeywords={and,not,set,in}, label=fig:TNES]
response = openai.chat.completions.create(
(*\fullbg{red!15}{-  model = "gpt-4o-2024-11-20", messages = messages)}*)
(*\fullbg{green!15}{+  model = "gpt-4o-2024-11-20", messages = messages,temperature = 1.0)}*)
\end{lstlisting}

%% file: Sections/Methodology.tex
Figure~\ref{fig:LLM_Methodo} presents our three-step methodology. 

\subsection{\textbf{Step 1}: Catalog Construction}
\label{sub:Step1}
\noindent\textbf{Goal:} Consolidate evidence from multiple sources into an \emph{LLM code smells catalog} with precise, implementation-oriented specifications.

\noindent\textbf{Inputs:}
i) Academic literature on LLM engineering and efficiency/reliability; 
ii) Grey literature (engineering blogs, provider documentations);  
iii) Empirical artifacts (GitHub, Stack Overflow (SO)). 

\noindent\textbf{Outputs:} 
A \emph{catalog of LLM code smells} documented using a common structure:
\textit{Name}, \textit{Context}, \textit{Problem}, \textit{Solution},  \textit{Example}, \textit{Sources}, \textit{Effect}.

\noindent\textbf{Procedure:}
We consolidate the final catalog in five steps, following established guidelines for systematic mapping studies in software engineering and multi-source triangulation~\cite{kitchenham2007guidelines}. (1) \emph{Search}: compose query families that pair LLM integration terms (e.g., “temperature”, “token budget”) with quality terms (e.g., “technical debt”, “defect”, “best practice”). 
(2) \emph{Screen}: title/abstract screening, then full-text reads; record evidence type and strength. 
(3) \emph{Synthesize}: normalize synonyms, merge duplicates, and draft code smell cards using a fixed schema. 
(4) \emph{Triangulate}: map each code smell to providers/APIs and cross-validate with their documentation and code examples. 
(5) \emph{Validate}: dual-review of each code smell, resolve disagreements by example-driven discussion.

A candidate code smell is admitted if it (i) is observable at the code level in LLM-integrating systems; (ii) has cross-source support ($>$5 of academic / grey / community / code change); and (iii) has an actionable remediation applicable across providers.

\subsection{\textbf{Step 2}: Detection Approach}
\label{sub:Step2}

\noindent\textbf{Goal:} Develop a detection tool for LLM code smells.

\noindent\textbf{Inputs:} The LLM code smell catalog from Step~\ref{sub:Step1}.

\noindent\textbf{Outputs:} A static detection tool: \chref{https://github.com/Brahim-Mahmoudi/SpecDetect4LLM_ICSE}{\textit{SpecDetect4LLM}}

\noindent\textbf{Procedure:} We extend \textit{SpecDetect4AI}~\cite{mahmoudi2025ai}, a static analyzer that detects AI-specific code smells through Domain Specific Language-defined rules. These rules allow the use of semantic predicates to detect coding patterns at a high level, avoiding any direct abstract syntax tree manipulation. Therefore, we:
(1) add LLM-specific detection rules derived from our catalog;
(2) introduce new reusable semantic predicates for common LLM integration idioms.
\textit{SpecDetect4LLM} is released as a versioned module of \textit{SpecDetect4AI} with comprehensive tests and documentation.
We select \textit{SpecDetect4AI} because it is an open-source tool that (i) demonstrated high accuracy against prior baselines, (ii) offers an extensible specification-driven design, and (iii) has strong reproducibility properties. 


\begin{figure}[ht]
    \centering
    \includegraphics[width=0.99\linewidth,height=155pt]{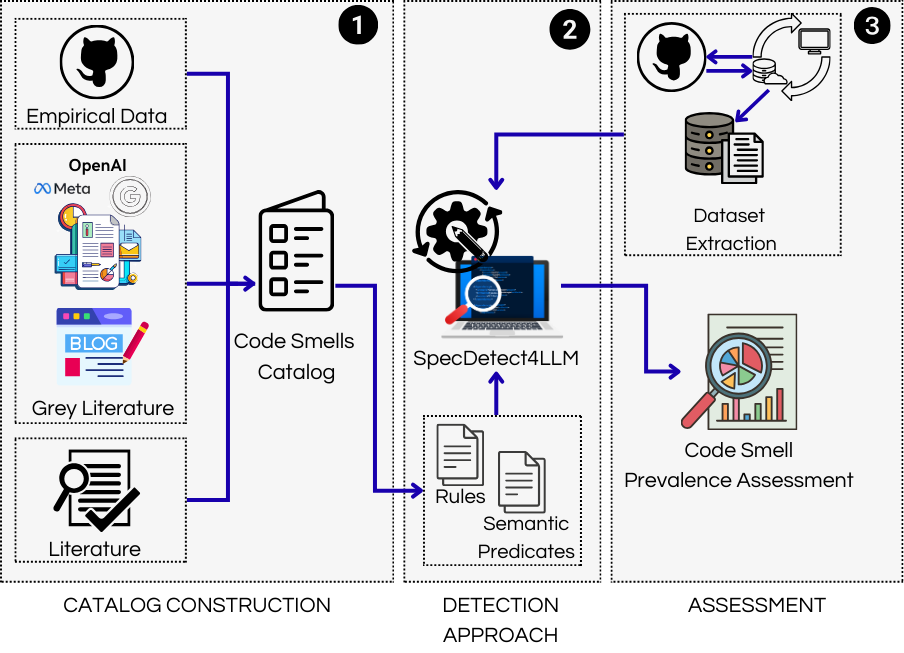}
    \caption{Workflow of the methodology}
    \label{fig:LLM_Methodo}
\end{figure}
\subsection{\textbf{Step 3}: Prevalence Assessment}
\label{sub:Step3}

\noindent\textbf{Goal:}  Run \chref{https://github.com/Brahim-Mahmoudi/SpecDetect4LLM_ICSE}{\textit{SpecDetect4LLM}} on a dataset of 200 LLM-integrating systems to deliver (i) per-smell prevalence and co-occurrence statistics and (ii) a precision-oriented manual audit demonstrating the practical relevance of the catalog.

\noindent\textbf{Inputs:} LLM-integrating Systems Dataset: a curated, versioned systems list with metadata.

\noindent\textbf{Outputs:} Prevalence Assessment: per-smell prevalence, density, and co-occurrence statistics, with breakdowns by framework, provider and application type.

\noindent\textbf{Procedure:}
Following the three-stage setup in Figure~\ref{fig:LLM_Methodo}, we (i) extract a dataset of open-source LLM-integrating systems, and (ii) assess the prevalence in the extracted dataset.

\textbf{i) \textit{Dataset Extraction}:}
We assemble a dataset of 200 LLM-integrating systems. First, we collect 100 open-source systems via the GitHub API, filtering by LLM keywords (e.g. \textit{openai}, \textit{llama}), programming language (Python), popularity (\textit{stars:>20}), and recency (sorted by last update). Then, we automatically verify LLM integration by parsing dependency files (e.g.,  \textit{requirements.txt}).

To strengthen external validity, we combine this collection with the 100 systems from Shao et al. \cite{shao2025llmscorrectlyintegratedsoftware}. This increases sample diversity as their set captures widely used and previously studied systems, while our crawl emphasizes more recent and actively maintained projects. We unify both sources, remove overlaps, and harmonize metadata (e.g. stars, last update). This construction yields a balanced dataset spanning established and emergent LLM-integrating systems, suitable for robust empirical validation.

\textbf{ii) \textit{Code Smell Prevalence Assessment}:} 
We execute our tool over the dataset under a standardized configuration, recording findings per file and per system. We then conduct a dual-review of a stratified sample of detections (reporting Cohen’s~$\kappa$) and audit false positives with targeted negative fixtures. For each code smell, we draw random samples until we reach at least 20 detections spanning at least 5 distinct systems. Each instance is labeled as \textit{true positive} (TP) or \textit{false positive} (FP) using a pre-defined decision sheet, enabling us to report per-smell precision $P=\frac{\text{TP}}{\text{TP}+\text{FP}}$. We did not estimate recall as it would require exhaustive file-level ground truthing and is therefore out-of-scope for this study. Likewise, the small precision sample size constrains the accuracy of our effectiveness assessment. Instead, we position our results as a precision-oriented prevalence study. All artifacts of this assessment are available in our replication package~\cite{Replication_package}.

%% file: Sections/Empirical_Validation.tex
We executed \textit{SpecDetect4LLM} \cite{Replication_package} on the full dataset of 200 systems and obtained 6,337 alerts across 60.50\% (121) of them.


\noindent \textbf{Observation 1-\textit{depth versus breadth}:}  
NMVP produces the most alerts (2,472) yet appears in fewer projects (72/200) than other smells. This pattern is consistent with alias propagation: once a moving tag (e.g., \textit{gpt-4o}) is adopted, it is repeated across many call sites inside the same projects, inflating counts.

\noindent \textbf{Observation 2-\textit{friction shapes prevalence}:}  
NSO and UMM have fewer raw alerts than NMVP, but affect the largest share of projects (NSO: 81/200, 40.50\%; UMM: 76/200, 38.00\%). Both require additional engineering work, schema design and strict parsing/validation for NSO and explicit limits on retries, timeouts, response size for UMM. Therefore, they surface early and persist across stacks. In the same vein, TNES is both frequent (1,374) and widespread (73/200), which matches the common practice of relying on framework defaults during prototyping and never revisiting them.

\noindent \textbf{Observation 3-\textit{prompt hygiene lags integration}:}
NSM is the least frequent (480) yet still present in more than half affected projects (69/121). Many teams wire up API calls before standardizing prompt roles, which weakens reproducibility and makes behavior sensitive to minor prompt edits.

\noindent \textbf{Reliability of findings:}  
A stratified manual audit shows high precision across the detection of all LLM code smells, for an average of 86.06\% (Table \ref{tab:defects-llm}). The imperfections in the detection are due to the static limits of \textit{SpecDetect4AI} \cite{mahmoudi2025ai}, which lacks support for dynamic detection and context-based analysis. Recall was not estimated, but the precision levels indicate that the observed prevalence patterns reflect real defects rather than noise.


\noindent \textbf{Implications:}  
Configuration and governance issues (NMVP, TNES) concentrate within projects as habits propagate, while engineering-effort code smells (NSO, UMM) diffuse broadly because removing them requires cross-cutting refactoring. Improving governance and reducing integration friction are likely to improve overall quality. 

%% file: Sections/Future_plans.tex
Our future research directions are clear. First, we plan to expand the catalog and further formalize our taxonomy by documenting additional LLM code smells and defining categories. This will be done through further research in the scientific literature and by analyzing common practices in open-source projects. Second, we will extend \chref{https://github.com/Brahim-Mahmoudi/SpecDetect4LLM_ICSE}{\textit{SpecDetect4LLM}} to include the newly documented smells and validate its detection effectiveness (precision and recall) through controlled experimentation. We will explore dynamic detection to capture execution-dependent effects beyond static detection. Finally, we will measure the impact of the identified code smells on the performance, robustness, reliability and maintainability of software systems through empirical studies. 


%% file: Sections/conclusion.tex
In this work, we introduced the concept of LLM code smells and presented a catalog of five, derived from relevant literature. We extended \textit{SpecDetect4AI} \cite{mahmoudi2025ai} to statically detect these smells and used it to  study their prevalence in 200 open-source systems. We found that all LLM code smells were present in 60.50\% of the studied projects, with a precision of 86.06\%. These findings confirm their relevance and the necessity of such a catalog and a dedicated detection tool. Overall, this work lays the foundation for providing clear and practical coding guidelines to improve code quality in LLM-integrating systems, and the necessary resources to support their adoption by practitioners.


%% file: bibli.bib
@article{MontandonEtAl2025DABC,
  author = {Jo{\~a}o Eduardo Montandon and Luciana Lourdes Silva and Cristiano Politowski and Daniel Prates and Arthur de Brito Bonif{\'a}cio and Ghizlane El Boussaidi},
  title = {Unboxing Default Argument Breaking Changes in Data Science Libraries},
  year = {2025},
  doi = {10.48550/arXiv.2408.05129},
  eprint = {2408.05129},
  archivePrefix = {arXiv},
  primaryClass = {cs.SE},
  url = {https://arxiv.org/abs/2408.05129},
  note = {JSS}
}

@inproceedings{MinhEtAl2025MinpSampling,
  author = {Nguyen Nhat Minh and Andrew Baker and Clement Neo and Allen G. Roush and Andreas Kirsch and Ravid Shwartz{-}Ziv},
  title = {Turning Up the Heat: Min-p Sampling for Creative and Coherent LLM Outputs},
  booktitle = {ICLR 2025},
  year = {2025}
}

@online{VellumTemperature2025,
  author = {{Vellum AI}},
  title = {LLM Temperature: How It Works and When You Should Use It},
  year = {2025},
  url = {https://www.vellum.ai/llm-parameters/temperature},
  urldate = {2025-09-25}
}

@online{OpenAIStructuredOutput,
  author = {{OpenAI}},
  title = {API Reference - Structured model outputs},
  year = {2025},
  url = {https://platform.openai.com/docs/guides/structured-outputs},
  urldate = {2025-09-25}
}

@online{OllamaModelfile2025,
  author = {{Ollama}},
  title = {Modelfile: Valid parameters and values},
  year = {2025},
  url = {https://github.com/ollama/ollama/blob/main/docs/modelfile.md#valid-parameters-and-values},
  urldate = {2025-09-25}
}

@online{PromptHub2025SystemMessages,
  author = {Cleary, Dan},
  title = {System Messages: Best Practices, Real-world Experiments \& Prompt Injection Protectors},
  year = {2025},
  month = {4},
  day = {29},
  url = {https://www.prompthub.us/blog/everything-system-messages-how-to-use-them-real-world-experiments-prompt-injection-protectors},
  urldate = {2025-09-25},
  note = {PromptHub Blog}
}

@online{StackOverflow2023SystemRoleUseCase,
  author = {{cyz3a5c0v1}},
  title = {What is the use case of System role},
  year = {2023},
  month = {5},
  day = {17},
  url = {https://stackoverflow.com/questions/76272624/what-is-the-use-case-of-system-role},
  urldate = {2025-09-25}
}

@article{WilsonEtAl2014BestPractices,
  author = {Wilson, Greg and Aruliah, D. A. and Brown, C. Titus and Chue Hong, Neil P. and Davis, Matt and Guy, Richard T. and Haddock, Steven H. D. and Huff, Kathryn D. and Mitchell, Ian M. and Plumbley, Mark D. and Waugh, Ben and White, Ethan P. and Wilson, Paul},
  title = {Best Practices for Scientific Computing},
  journal = {PLOS Biology},
  year = {2014},
  volume = {12},
  number = {1},
  pages = {e1001745}
}

@article{VenturiniEtAl2023IDepended,
  author = {Venturini, Daniel and Cogo, Filipe Roseiro and Polato, Ivanilton and Gerosa, Marco A. and Wiese, Igor Scaliante},
  title = {I Depended on You and You Broke Me: An Empirical Study of Manifesting Breaking Changes in Client Packages},
  year = {2023},
  doi = {10.48550/arXiv.2301.04563},
  eprint = {2301.04563},
  archivePrefix = {arXiv},
  url = {https://arxiv.org/abs/2301.04563},
  note = {TOSEM, 2023}
}

@inproceedings{ReyesEtAl2024BUMP,
  author = {Reyes, Frank and Gamage, Yogya and Skoglund, Gabriel and Baudry, Benoit and Monperrus, Martin},
  title = {BUMP: A Benchmark of Reproducible Breaking Dependency Updates},
  booktitle = {SANER 2024},
  year = {2024},
  doi = {10.48550/arXiv.2401.09906},
  eprint = {2401.09906},
  archivePrefix = {arXiv},
  url = {https://arxiv.org/abs/2401.09906},
  urldate = {2025-09-25}
}

@online{Microsoft2025FoundationLifecycle,
  author = {{Microsoft Learn}},
  title = {Design to Support Foundation Model Life Cycles},
  year = {2025},
  month = {5},
  day = {16},
  url = {https://learn.microsoft.com/en-us/azure/architecture/ai-ml/guide/manage-foundation-models-lifecycle},
  urldate = {2025-09-25}
}

@online{OpenRouterHome2025,
  author = {{OpenRouter}},
  title = {OpenRouter.ai - One API for Any Model},
  year = {2025},
  url = {https://openrouter.ai/},
  urldate = {2025-09-25}
}

@online{Kharitonov2024EnforcingJSON,
  author = {Kharitonov, Daniel},
  title = {Enforcing JSON Outputs in Commercial LLMs},
  year = {2024},
  month = {8},
  url = {https://medium.com/data-science/enforcing-json-outputs-in-commercial-llms-3db590b9b3c8},
  urldate = {2025-09-25},
  note = {Towards Data Science}
}

@online{DeveloperService2025PydanticLLM,
  author = {{Developer Service}},
  title = {A Practical Guide on Structuring LLM Outputs with Pydantic},
  year = {2025},
  month = {6},
  day = {12},
  url = {https://dev.to/devasservice/a-practical-guide-on-structuring-llm-outputs-with-pydantic-50b4},
  urldate = {2025-09-25}
}

@online{WymanBarber2024ValidateOutputs,
  author = {Wyman, Matt and Barber, Sarah},
  title = {How to Validate the Output of LLM-Based Products},
  year = {2024},
  month = {7},
  day = {20},
  url = {https://okareo.com/blog/posts/validate-llm-output},
  urldate = {2025-09-25}
}

@online{Modelmetry2024JSONSchema,
  author = {{Modelmetry}},
  title = {How To Ensure LLM Output Adheres to a JSON Schema},
  year = {2024},
  month = {10},
  day = {28},
  url = {https://modelmetry.com/blog/how-to-ensure-llm-output-adheres-to-a-json-schema},
  urldate = {2025-09-25}
}

@online{LangChain2023Issue3709,
  author = {{mariafilippa}},
  title = {PydanticOutputParser has high chance failing when completion contains new line \#3709},
  year = {2023},
  month = {4},
  day = {28},
  url = {https://github.com/hwchase17/langchain/issues/3709},
  urldate = {2025-09-25},
  note = {GitHub issue}
}

@software{ArenaAI2025StructuredLogprobs,
  author = {{arena-ai}},
  title = {structured-logprobs},
  year = {2025},
  url = {https://github.com/arena-ai/structured-logprobs},
  urldate = {2025-09-25},
  note = {OS library: enhances OpenAI Structured Outputs with token logprobs}
}

@inproceedings{HanEtAl2025TokenBudgetAware,
  author = {Han, Tingxu and Wang, Zhenting and Fang, Chunrong and Zhao, Shunyuan and Ma, Shiqing and Chen, Ziang},
  title = {Token-Budget-Aware LLM Reasoning},
  booktitle = {ACL 2025},
  year = {2025},
  pages = {24842--24855},
  publisher = {Association for Computational Linguistics},
  doi = {10.18653/v1/2025.findings-acl.1274},
  url = {https://aclanthology.org/2025.findings-acl.1274/}
}

@online{BigQuery2025Quotas,
  author = {{Google Cloud}},
  title = {Quotas and limits - BigQuery},
  year = {2025},
  url = {https://cloud.google.com/bigquery/quotas},
  urldate = {2025-09-25}
}

@book{10.5555/311424,
  author = {Martin Fowler and Kent Beck and John Brant and William Opdyke and Don Roberts},
  title = {Refactoring: improving the design of existing code},
  year = {1999},
  publisher = {Addison-Wesley Longman Publishing Co., Inc.},
  address = {USA},
  isbn = {0201485672}
}

@article{Carvalho2019AndroidSmells,
  author = {Carvalho, Silvio G. and Aniche, Maur{\'i}cio and Ver{\'i}ssimo, Jo{\~a}o and Garcia, Alessandro and Alves, Vitor and Gheyi, Rohit},
  title = {An empirical catalog of code smells for the presentation layer of Android apps},
  journal = {Empirical Software Engineering},
  year = {2019},
  volume = {24},
  number = {6},
  pages = {3546--3586},
  publisher = {Springer},
  doi = {10.1007/s10664-019-09768-9},
  url = {https://doi.org/10.1007/s10664-019-09768-9},
  month = {dec}
}

@article{zhang2022codesmellsmachinelearning,
  author = {Haiyin Zhang and Luís Cruz and Arie van Deursen},
  title = {Code Smells for ML Applications},
  year = {2022},
  eprint = {2203.13746},
  archivePrefix = {arXiv},
  primaryClass = {cs.SE},
  url = {https://arxiv.org/abs/2203.13746}
}

@article{ning2024definingdetectingdefectslarge,
  author = {Kaiwen Ning and Jiachi Chen and Jingwen Zhang and Wei Li and Zexu Wang and Yuming Feng and Weizhe Zhang and Zibin Zheng},
  title = {Defining and Detecting the Defects of the Large Language Model-based Autonomous Agents},
  year = {2024},
  eprint = {2412.18371},
  archivePrefix = {arXiv},
  primaryClass = {cs.SE},
  url = {https://arxiv.org/abs/2412.18371}
}

@article{ke2025surveyfrontiersllmreasoning,
  author = {Zixuan Ke and Fangkai Jiao and Yifei Ming and Xuan-Phi Nguyen and Austin Xu and Do Xuan Long and Minzhi Li and Chengwei Qin and Peifeng Wang and Silvio Savarese and Caiming Xiong and Shafiq Joty},
  title = {A Survey of Frontiers in LLM Reasoning: Inference Scaling, Learning to Reason, and Agentic Systems},
  year = {2025},
  eprint = {2504.09037},
  archivePrefix = {arXiv},
  primaryClass = {cs.AI},
  url = {https://arxiv.org/abs/2504.09037}
}

@article{shao2025llmscorrectlyintegratedsoftware,
  author = {Yuchen Shao and Yuheng Huang and Jiawei Shen and Lei Ma and Ting Su and Chengcheng Wan},
  title = {Are LLMs Correctly Integrated into Software Systems?},
  year = {2025},
  eprint = {2407.05138},
  archivePrefix = {arXiv},
  primaryClass = {cs.SE},
  url = {https://arxiv.org/abs/2407.05138}
}

@inproceedings{11081716,
  author = {Winston, Cailin and Just, René},
  title = {A Taxonomy of Failures in Tool-Augmented LLMs},
  booktitle = {AST 2025},
  year = {2025},
  pages = {125--135},
  doi = {10.1109/AST66626.2025.00019}
}

@article{cemri2025multiagentllmsystemsfail,
  author = {Mert Cemri and Melissa Z. Pan and Shuyi Yang and Lakshya A. Agrawal and Bhavya Chopra and Rishabh Tiwari and Kurt Keutzer and Aditya Parameswaran and Dan Klein and Kannan Ramchandran and Matei Zaharia and Joseph E. Gonzalez and Ion Stoica},
  title = {Why Do Multi-Agent LLM Systems Fail?},
  year = {2025},
  eprint = {2503.13657},
  archivePrefix = {arXiv},
  primaryClass = {cs.AI},
  url = {https://arxiv.org/abs/2503.13657}
}

@inproceedings{le-jeune-etal-2025-realharm,
  author = {Le Jeune, Pierre and Liu, Jiaen and Rossi, Luca and Dora, Matteo},
  title = {RealHarm: A Collection of Real-World Language Model Application Failures},
  booktitle = {LLMSEC 2025},
  year = {2025},
  month = {aug},
  pages = {87--100}
}

@article{ronanki2024promptsmellsomenundesirable,
  author = {Krishna Ronanki and Beatriz Cabrero-Daniel and Christian Berger},
  title = {Prompt Smells: An Omen for Undesirable Generative AI Outputs},
  year = {2024},
  eprint = {2401.12611},
  archivePrefix = {arXiv},
  primaryClass = {cs.LG},
  url = {https://arxiv.org/abs/2401.12611}
}

@article{tian2025taxonomypromptdefectsllm,
  author = {Haoye Tian and Chong Wang and BoYang Yang and Lyuye Zhang and Yang Liu},
  title = {A Taxonomy of Prompt Defects in LLM Systems},
  year = {2025},
  eprint = {2509.14404},
  archivePrefix = {arXiv},
  primaryClass = {cs.SE},
  url = {https://arxiv.org/abs/2509.14404}
}

@article{agrawal2025evaluatingperformancellminference,
  author = {Amey Agrawal and Nitin Kedia and Anmol Agarwal and Jayashree Mohan and Nipun Kwatra and Souvik Kundu and Ramachandran Ramjee and Alexey Tumanov},
  title = {On Evaluating Performance of LLM Inference Serving Systems},
  year = {2025},
  eprint = {2507.09019},
  archivePrefix = {arXiv},
  primaryClass = {cs.LG},
  url = {https://arxiv.org/abs/2507.09019}
}

@article{zhuo2025identifyingmitigatingapimisuse,
  author = {Terry Yue Zhuo and Junda He and Jiamou Sun and Zhenchang Xing and David Lo and John Grundy and Xiaoning Du},
  title = {Identifying and Mitigating API Misuse in Large Language Models},
  year = {2025},
  eprint = {2503.22821},
  archivePrefix = {arXiv},
  primaryClass = {cs.SE},
  url = {https://arxiv.org/abs/2503.22821}
}

@article{esfahani2024understandingdefectsgeneratedcodes,
  author = {Ali Mohammadi Esfahani and Nafiseh Kahani and Samuel A. Ajila},
  title = {Understanding Defects in Generated Codes by Language Models},
  year = {2024},
  eprint = {2408.13372},
  archivePrefix = {arXiv},
  primaryClass = {cs.SE},
  url = {https://arxiv.org/abs/2408.13372}
}

@article{mahmoudi2025ai,
  author = {Mahmoudi, Brahim and Moha, Naouel and Stievenert, Quentin and Avellaneda, Florent},
  title = {AI-Specific Code Smells: From Specification to Detection},
  year = {2025},
  doi = {10.48550/arXiv.2509.20491},
  eprint = {2509.20491},
  archivePrefix = {arXiv},
  primaryClass = {cs.SE},
  url = {https://arxiv.org/abs/2509.20491}
}

@techreport{kitchenham2007guidelines,
  author = {Kitchenham, Barbara and Charters, Stuart},
  title = {Guidelines for performing Systematic Literature Reviews in Software Engineering},
  institution = {EBSE 2007},
  year = {2007},
  number = {EBSE-2007-01},
  url = {https://www.elsevier.com/__data/promis_misc/525444systematicreviewsguide.pdf}
}

@online{Replication_package,
  author = {Brahim, Mahmoudi and Zacharie, Chenail-larcher},
  title = {Replication\_Package\_LLM\_code\_smells},
  year = {2025},
  url = {https://github.com/Brahim-Mahmoudi/SpecDetect4LLM_ICSE}
}

@article{xia2025analyzing16193llmpapers,
  author = {Zhiqiu Xia and Lang Zhu and Bingzhe Li and Feng Chen and Qiannan Li and Chunhua Liao and Feiyi Wang and Hang Liu},
  title = {Analyzing 16,193 LLM Papers for Fun and Profits},
  year = {2025},
  eprint = {2504.08619},
  archivePrefix = {arXiv},
  primaryClass = {cs.DL},
  url = {https://arxiv.org/abs/2504.08619}
}

@phdthesis{khatun2024reliability,
  author = {Khatun, Aisha},
  title = {Uncovering the Reliability and Consistency of AI Language Models: A Systematic Study},
  school = {University of Waterloo},
  year = {2024},
  month = {August},
  url = {https://uwspace.uwaterloo.ca/items/e01e11a6-e033-4f6a-85c6-849fba74e039}
}

@article{YANG2025113503,
  author = {Wenli Yang and Lilian Some and Michael Bain and Byeong Kang},
  title = {A comprehensive survey on integrating large language models with knowledge-based methods},
  journal = {Knowledge-Based Systems},
  year = {2025},
  volume = {318},
  pages = {113503},
  doi = {https://doi.org/10.1016/j.knosys.2025.113503},
  issn = {0950-7051}
}

@article{bucaioni2025functionalsoftwarereferencearchitecture,
  author = {Alessio Bucaioni and Martin Weyssow and Junda He and Yunbo Lyu and David Lo},
  title = {A Functional Software Reference Architecture for LLM-Integrated Systems},
  year = {2025},
  eprint = {2501.12904},
  archivePrefix = {arXiv},
  primaryClass = {cs.SE},
  url = {https://arxiv.org/abs/2501.12904}
}

@online{OpenAIDocs2025,
  author = {{OpenAI}},
  title = {OpenAI Platform Documentation},
  year = {2025},
  url = {https://platform.openai.com/docs},
  urldate = {2025-09-25}
}

@online{AnthropicDocs2025,
  author = {{Anthropic}},
  title = {Claude Documentation (API, Models)},
  year = {2025},
  url = {https://docs.claude.com/},
  urldate = {2025-09-25}
}

@online{GoogleGeminiDocs2025,
  author = {{Google}},
  title = {Gemini API - Google AI for Developers},
  year = {2025},
  url = {https://ai.google.dev/},
  urldate = {2025-09-25}
}

@online{AzureOpenAIDocs2025,
  author = {{Microsoft Learn}},
  title = {Azure OpenAI - Documentation (Quotas, Structured Outputs, How-To)},
  year = {2025},
  url = {https://learn.microsoft.com/en-us/azure/ai-foundry/openai/},
  urldate = {2025-09-25}
}

@online{HFTransformersDocs2025,
  author = {{Hugging Face}},
  title = {Transformers Documentation (Generation, Chat Templates, Model Revisions)},
  year = {2025},
  url = {https://huggingface.co/docs/transformers},
  urldate = {2025-09-25}
}

@article{Morishige/EnsureReprod,
  author = {Morishige, Masumi and Koshihara, Ryo},
  title = {Ensuring Reproducibility in Generative AI Systems for General Use Cases: A Framework for Regression Testing and Open Datasets},
  year = {2025},
  doi = {10.48550/arXiv.2505.02854},
  eprint = {2505.02854},
  archivePrefix = {arXiv},
  url = {https://arxiv.org/abs/2505.02854},
  month = {5}
}

@inproceedings{NeedStructuredOutput/10.1145/3613905.3650756,
  author = {Liu, Michael Xieyang and Liu, Frederick and Fiannaca, Alexander J. and Koo, Terry and Dixon, Lucas and Terry, Michael and Cai, Carrie J.},
  title = {We Need Structured Output: Towards User-centered Constraints on Large Language Model Output},
  booktitle = {Extended Abstracts of the CHI Conference on Human Factors in Computing Systems},
  year = {2024},
  publisher = {Association for Computing Machinery},
  doi = {10.1145/3613905.3650756},
  url = {https://doi.org/10.1145/3613905.3650756},
  isbn = {9798400703317},
  address = {New York, NY, USA}
}

@article{jeong2025messagegenerationuserpreferences,
  author = {Minbyul Jeong and Jungho Cho and Minsoo Khang and Dawoon Jung and Teakgyu Hong},
  title = {System Message Generation for User Preferences using Open-Source Models},
  year = {2025},
  eprint = {2502.11330},
  archivePrefix = {arXiv},
  primaryClass = {cs.CL},
  url = {https://arxiv.org/abs/2502.11330}
}

@inproceedings{Neumann_2025,
  author = {Neumann, Anna and Kirsten, Elisabeth and Zafar, Muhammad Bilal and Singh, Jatinder},
  title = {Position is Power: System Prompts as a Mechanism of Bias in Large Language Models (LLMs)},
  booktitle = {Proceedings of the 2025 ACM Conference on Fairness, Accountability, and Transparency},
  year = {2025},
  month = {jun},
  pages = {573--598},
  publisher = {ACM},
  doi = {10.1145/3715275.3732038},
  url = {http://dx.doi.org/10.1145/3715275.3732038}
}

@online{aws2025timeouts,
  author = {{Amazon Web Services}},
  title = {Timeouts, retries and backoff},
  year = {2025},
  url = {https://aws.amazon.com/fr/builders-library/timeouts-retries-and-backoff-with-jitter},
  urldate = {2025-09-25}
}

@article{chen2025adaptivelyrobustllminference,
  author = {Zixi Chen and Yinyu Ye and Zijie Zhou},
  title = {Adaptively Robust LLM Inference Optimization under Prediction Uncertainty},
  year = {2025},
  eprint = {2508.14544},
  archivePrefix = {arXiv},
  primaryClass = {cs.LG},
  url = {https://arxiv.org/abs/2508.14544}
}

@software{openai_python2025,
  author = {{OpenAI}},
  title = {openai-python},
  year = {2025},
  url = {https://github.com/openai/openai-python},
  urldate = {2025-09-25}
}
